\documentclass[11pt]{article}
\title{Quantum mechanics and the Continuum Problem (II)}
\author{O. Yaremchuk\thanks{e-mail: yarem@sci.lebedev.ru}}
\date{\today}
\begin{document}
\maketitle
\begin{abstract}

In one-dimensional case, it is shown that the basic principles
of quantum mechanics are properties of the set of intermediate
cardinality.

\end{abstract}

The concept of discrete space is not a unique alternative
of the continuous space. Since discrete space is a countable set,
there is an intermediate possibility connected with the continuum
problem: space may be neither continuous nor discrete.
The commonly held view is that the independence of the
continuum hypothesis (CH) is not a certain solution of the
continuum problem in consequence of incompleteness of set theory.
Nevertheless, from the independence of CH follows a unique
definite status of the set of intermediate cardinality.
It is important here that this set must be a subset of continuum
(continuum must contain a subset equivalent to the intermediate set).
Taking into account that any separation of the subset is a proof of
existence of the intermediate set, which contradicts the independence
of CH, we get that the set of intermediate cardinality exists only as
a subset of continuum. In other words, the subset of intermediate
cardinality, in principle, cannot be separated from continuum
(set theory ``confinement''). If Zermelo-Fraenkel set theory is
consistent, complete, and giving the correct description of the
notion of set, then this is the only possible understanding of the
independence of CH.

Note that if we postulate existence of the intermediate set (in other
words, if we take the negation of CH as an axiom), the result will be the
same: since any construction or separation of the set are forbidden by the
independence of CH, we have to reconcile with the same ``latent''
intermediate subset in continuum which we can get without any additonal
assumption. And it is not reasonable to take CH as an axiom because, as a
consequence, we lose this subset.

According to the separation axiom schema, for any set
$X$ and for any property expressed by formula $\varphi$ 
there exists a subset of the set $X$, which contains
only members of $X$ having $\varphi$. Then some subset
cannot be separated from continuum if each point of the
subset does not have its own peculiar properties
but only combines properties of the members of the
countable set and continuum.

At first sight, this seems to be meaningless.
But the content of the requirement coincides with the
content of wave-particle duality: quantum particle
combines properties of a wave (continuum) and a
point-like particle (the countable set).

As an illustration, consider a brick road which consists of
black bricks and white bricks. If we know (or suspect) that
among them there are some bricks which have white top
side and black bottom side (or vice versa), we, nevertheless,
cannot find them. Based only on top view, the problem of
separation (and even existence) of black-and-white bricks is
undecidable. Each brick can be black-and-white with some
probability. However, if we have top view and bottom view,
we can find these bricks: each of them looks like a white
brick on the one view and like a black brick on the other view
(``black-white duality''). 

In order to get information about the ``invisible'' set
consider the maps of the intermediate set $I$ to
the sets of real numbers ($R$) and natural numbers ($N$).

Let the map $I\to N$ decompose $I$ into
the countable set of equivalent mutually disjoint
infinite subsets: $\cup I_n=I$ ($n\in N$).
Let $I_n$ be called a unit set. All members of $I_n$
have the same countable coordinate $n$.

Consider the map $I\to R$.
Continuum $R$ contains a subset $M$ equivalent to $I$,
i.e., there exists a bijection
\begin{equation}
f:I\to M\subset R.
\end{equation}
This bijection reduces to a separation of the intermediate
subset $M$ from continuum. Since any separation
procedure is a proof of existence of the intermediate
set and, therefore, contradicts the independence
of the continuum hypothesis, we, in principle, do not
have a rule for assigning a definite real number to a
point of the intermediate set. Hence, any bijection can
take a point of the intermediate set only to a random
real number. If we do not have preferable real numbers,
then we have the equiprobable mapping. This already
conforms to the quantum free particle. In the general
case, we have the probability $P(r)dr$ of finding a
point $s\in I$ about $r$.

Thus the point of the intermediate set has two coordinates:
a definite natural number and a random real number:
\begin{equation}\label{s}
s:(n,r_{random}).
\end{equation}
Only the natural number coordinate gives reliable
information about the relative positions of the points
of the set and the size of its interval. 
But the points of a unit set are indistinguishable. It is
clear that the probability $P(r)$ depends on the
natural number coordinate of the corresponding point.
Note that the information about a point in the
one-dimensional intermediate set is necessarily
two-dimensional.

For two real numbers $a$ and $b$ the probability
$P_{a\cup b}dr$ of finding $s$ in the union of the
neighborhoods $(dr)_a\cup (dr)_b$
\begin{equation}
P_{a\cup b}\,dr\ne [P(a)+P(b)]\,dr
\end{equation}
because $s$ corresponds to both (all) points at the
same time (the events are not mutually exclusive).
It is convenient to introduce a function $\psi(r)$
such that $P(r)={\cal P}[\psi(r)]$ and
$\psi_{a\cup b}=\psi(a)+\psi(b)$.
The idea is to compute the non-additive probability
from some additive object by a simple rule.

We have
\begin{equation}
P_{a\cup b}={\cal P}(\psi_{a\cup b})={\cal P}[\psi(a)+\psi(b)]\ne {\cal P}[\psi(a)]+{\cal P}[\psi(b)],
\end{equation}
i.e., the dependence ${\cal P}[\psi (r)]$ is non-linear.
The simplest non-linear dependence is a square
dependence:
\begin{equation}
{\cal P}[\psi(r)]=|\psi(r)|^2.
\end{equation} \label{born}

The probability $P(r)$ is not probability density
because we cannot integrate it due to its non-additivity
(an integral is a sum). The normalization condition
means only that $f$ is a bijection: we can find only one
image of the point $s$ in $R$.
Actually, the concept of probability should be modified.
An illustration in terms of the above brick road will make
this clear: If we know the exact number $N_{B-W}$ of the
black-and-white bricks, we do not need to check all
the bricks of perhaps infinite brick road. It is reasonable
to stop checking when all this bricks are obtained and
put 
\begin{equation}
P_{B-W}=\frac{N_{B-W}}{N_{checked}},
\end{equation}
where $P_{B-W}$ is the probability of finding a black-and-white
brick, $N_{checked}$ is the exact (minimal) number of the bricks
checked. Thus only $N_{checked}$ may vary in the different test runs
(finding  all the black-and-white bricks) and we have to use the
average value.

The concept of probability for continuum may be modified in a similar
way, since the point always may be found in a finite interval.
We do not need to take into consideration remaining empty continuum.

But we shall not alter the concept of probability because it
is not altered in quantum mechanics (although this results in
infinite probabilities). The main purpose of this paper is to
show that quantum mechanics describes the set of intermediate
cardinality.

The function $\psi$, necessarily, depends on $n$: $\psi(r)\to\psi(n,r)$.
Since $n$ is accurate up to a constant (shift)
and the function $\psi$ is defined up to the 
factor $e^{i\mbox{const}}$, we have
\begin{equation}
\psi (n+\mbox{const},r)=e^{i\mbox{const}}\psi (n,r).
\end{equation}
Hence, the function $\psi$ is of the following form:
\begin{equation}
\psi (r,n)=A(r)e^{2\pi in}\label{debr}.
\end{equation}

Thus the point of the intermediate set corresponds to
the function Eq.(\ref{debr}) in continuum. We can specify
the point by the function $\psi(n,r)$ before the mapping
and by the random real number and the natural
number when the mapping has performed. In other words,
the function $\psi(n,r)$ may be regarded as the image
of $s$ in $R$ between mappings.

Consider probability $P(a,b)$ of finding the point $s$ at $b$
after finding it at $a$.
Let us use a continuous parameter $t$ for correlation
between continuous and countable coordinates of the point
$s$ (simultaneity) and in order to distinguish between
the different mappings (events ordering):
\begin{equation}
r(t_a),n(t_a)\to\psi(t)\to r(t_b),n(t_b),
\end{equation}\label{0-t}
where $t_a<t<t_b$ and $\psi(t)=\psi[n(t),r(t)]$.
For simplicity, we shall identify the parameter with
time without further discussion. Note that we cannot use the direct
dependence $n=n(r)$. Since $r=r(n)$ is a random number,
the inverse function is meaningless.

Assume that $s$ is a ``observable'' point, i.e., for
each $t\in (t_a,t_b)$ there exists the image of the point
in continuum $R$. 

Partition interval $(t_a,t_b)$ into $k$ equal parts
$\varepsilon$:
\begin{eqnarray}
k\varepsilon =t_b-t_a,\nonumber\\
\varepsilon =t_i-t_{i-1},\nonumber\\
t_a=t_0,t_b=t_k,\\
a=r(t_a)=r_0,\, b=r(t_k)=r_k.\nonumber
\end{eqnarray}\label{partition}
The conditional probability of of finding the point $s$ at
$r(t_i)$ after $r(t_{i-1})$ is given by
\begin{equation}\label{cond}
P(r_{i-1},r_i)=\frac{P(r_i)}{P(r_{i-1})}
\end{equation}
(between the points $t_{i-1}$ and $t_i$, the continuous image of
the point is out of control but the unmonitored zone will be reduced
to zero by passage to the limit $\varepsilon\to 0$),
i.e.,
\begin{equation}
P(r_{i-1},r_i)=\left|\frac{A_i}{A_{i-1}}e^{2\pi i\Delta n_i}\right|^2,
\end{equation}
where $\Delta n_i=|n(t_i)-n(t_{i-1})|$. Note that $\Delta n_i$ is
really a vector.

The probability of the sequence of the transitions (we may use the word
``transition'' because we have the substantiated notion of time)
\begin{equation}
r_0,\ldots ,r_i,\ldots r_k
\end{equation}\label{sequence}
is given by
\begin{equation}
P(r_0,\ldots ,r_i,\ldots r_k)=P(r_1,r_2)\cdots P(r_{i-1},r_i)\cdots P(r_{k-1},r_k),
\end{equation}
i.e.,
\begin{equation}
P(r_0,\ldots ,r_i,\ldots r_k)=\left|\frac{A_k}{A_0}\exp 2\pi i\sum_{i=1}^k\Delta n_i\right|^2.
\end{equation}
Then probability of the corresponding continuous sequence
of the transitions $r(t)$
\begin{equation}\label{pathprob}
P[r(t)]=\lim_{\varepsilon\to 0}P(r_0,\ldots ,r_i,\ldots r_k)=\left|\frac{A_k}{A_0}e^{2\pi im}\right|^2,
\end{equation}
where
\begin{equation}
m=\lim_{\varepsilon\to 0}\sum_{i=1}^k\Delta n_i.
\end{equation}

Since at any time $t_a<t<t_b$ the point $s$ corresponds to all
points of $R$, it also corresponds to all continuous random sequences of
mappings $r(t)$ simultaneously (we emphasize that $r(t)$ is not
necessarily a classical path).

Probability $P[r(t)]$ of finding the point at any time
$t_a\leq t\leq t_b$ on $r(t)$ is non-additive too.
Therefore, we introduce an additive functional $\phi[r(t)]$.
In the same way as above, we get
\begin{equation}
P[r(t)]=|\phi[r(t)]|^2.
\end{equation}\label{phi}

Taking into account Eq.(\ref{pathprob}), we can put
\begin{equation}
\phi[r(t)]=\frac{A_N}{A_0}e^{2\pi im}=\mbox{const}\,e^{2\pi im}.
\end{equation}
Thus we have
\begin{equation}
P(a,b) = |\!\!\sum_{all\,r(t)}\!\!\mbox{const}\,e^{2\pi im}|^2,
\end{equation}
i.e., the probability $P(a,b)$ of finding the point $s$ at $b$
after finding it at $a$ satisfies the conditions of Feynman's approach
(section 2-2 of \cite{Feynman}) for $S/\hbar=2\pi m$ (indeed, Feynman
does not essentially use in Chap. 2 that $S/\hbar$ is just action).

Therefore,
\begin{equation}
P(a,b)=|K(a,b)|^2,
\end{equation}
where $K(a,b)$ is path integral (2-25) of \cite{Feynman}:
\begin{equation}\label{pathint}
K(a,b)=\int_{r_a}^{r_b}\!e^{2\pi im}D r(t).
\end{equation}

Thus we can apply Feynman's method in the following way.

1)We substitute $2\pi m$ for $S/\hbar$ in in Eq.(2-15) of \cite{Feynman}.

2)In section 2-3 of \cite{Feynman} Feynman explains how
the principle of least action follows from the dependence
\begin{equation}\label{sum}
P(a,b)= |\!\sum_{all\,r(t)}\!\!\mbox{const}\,e^{(i/\hbar)S[r(t)]}|^2.
\end{equation}
By the same nonrigourous reasoning, for ``very, very'' large $m$,
we get ``the principle of least $m$''.
This also means that for large $m$ the point $s$ has a definite
stationary path and, consequently, a definite continuous coordinate.
In other words, the corresponding interval of the intermediate
set is sufficiently close to continuum (let the interval be
called macroscopic), i.e., cardinality of the intermediate set depends
on its size. Recall that we can measure the size of an interval of
the set only in the unit sets (some packets of points).

3)Since large $m$ and $\Delta n_i$ may be considered as continuous
variables, we have
\begin{equation}\label{lim}
m=\lim\limits_{\varepsilon\to 0}\sum_{i=1}^N\Delta n_i=\int_{t_a}^{t_b}\!\!dn(t)=min.
\end{equation}
The function $n(t)$ may be regarded as some function of
$r(t)$: $n(t)=\eta[r(t)]$. It is important that $r(t)$ is not
random due to the second item. Therefore,
\begin{equation}\label{f}
\int_{t_a}^{t_b}\!\!dn(t)=\int_{t_a}^{t_b}\!\frac{d\eta}{dr}\,\dot{r}\,dt=min,
\end{equation}
where $\frac{d\eta}{dr}\,\dot{r}$ is some function of $r$, $\dot{r}$, and $t$
(note absence of higher time derivatives than $\dot{r}$), i.e.,
large $m$ can be identified with action:
\begin{equation}
m=\int_{t_a}^{t_b}\!\! L(r,\dot{r},t)\,dt=min.
\end{equation}
Since the value of action depends on units of measurement, we need
a parameter $h$ (depending on units only) such that
\begin{equation}
hm=\int_{t_a}^{t_b}\!\! L(r,\dot{r},t)\,dt.
\end{equation}
Note that we can substitute action for $m$ only for sufficiently
high time rate of change of the countable coordinate $n$ because,
if $\Delta n_i=n(t_{i})-n(t_{i-1})$ in Eq.(\ref{lim}) is not
sufficiently large to be considered as an (even infinitesimal)
interval of continuum, action reduces to zero. This may be understood
as vanishing of mass of the point.
Recall that mass is a factor which appear in
Lagrangian of a free point as a peculiar property of the point
under consideration, i.e., formally, mass may be regarded as a
consequence of the principle of least action \cite{mech}.

Finally, we may substitute $S/\hbar$ for $2\pi m$ in Eq.(\ref{pathint})
and apply Feynman's method to the set of intermediate
cardinality.

Consider the special case of constant time rate of change $\nu$ of
the countable coordinate $n$. We have $m=\nu(t_b-t_a)$. Then
``the principle of least $m$'' reduces to
``the principle of least $t_b-t_a$''.
If $\nu$ is not sufficiently large (massless point), this is the
simplest form of Fermat's least time principle for light. The more
general form of Fermat's principle follows from Eq.(\ref{lim}): since
\begin{equation}
\int_{t_a}^{t_b}\!\!dn(t)=\nu\!\int_{t_a}^{t_b}\!\!dt=min,
\end{equation}
we obviously get
\begin{equation}\label{fermat}
\int_{t_a}^{t_b}\!\!\frac{dr}{v(t)}=min,
\end{equation}
where $v(t)=dr/dt$.
In the case of non-zero action (mass point), the principle
of least action and Fermat's principle ``work''
simultaneously. It is clear that any additional factor
can only increase the ``pure least'' time. As a result
$t_b-t_a$ for a massless point bounds below $t_b-t_a$
for any other point and, therefore, $(b-a)/(t_b-t_a)$
for massless point bounds above average speed between 
the same points $a$ and $b$ for continuous image of any
point of the intermediate set.
This is a step towards special relativity.

It is important to make some general remarks on the
description of the set intermediate cardinality.

The complete description of the intermediate set falls
into two basic parts: continuous and countable. 
The continuous description is classical mechanics
(the principle of least action is an intrinsic
property of the set of intermediate cardinality).

Quantum mechanics is a connecting link and must be
considered as a separate description (a countable
description in terms of the continuous one). The
description has its particular transitional main law
(with action but without the principle of least action):
the wave equation. Therefore, quantum mechanics is
relevant for sufficiently large interval which may be
considered as continuum. Compare this with the Copenhagen
macroscopic measuring apparatus. 

Thus the complete description of the intermediate set
consist of three parts:
macroscopic (continuous), microscopic in macroscopic
terms (let us call it ``submicroscopic''), and proper
microscopic, i. e., it is a system of three dual theories.

Mathematical ``invisibility'' of the intermediate set
leads to confusion: all descriptions are placed in the
same continuous space. As a result the directions of the
countable descriptions are lost and replaced with spin.
We also lose microscopic dimensions of non-continuous
descriptions.

The total number of space time dimensions of
three 3D descriptions is ten. The same number of
dimensions appear in string theories. But the extra
dimensions of the intermediate set are essentially
microscopic and do not require compactification.
Since microscopic intervals (unlike macroscopic ones)
are essentially non-equivalent, the proper microscopic
description must split into a system of countable (quantum)
dual ``theories'' with number of extra dimensions
corresponding to the number of distinguishable cardinalities.

By definition, a proper microscopic interval can not be
considered as continuous, i.e., it has no length. In other
words, its macroscopic (continuous) image is exactly a point.
Thus from macroscopic point of view, there are two kinds
of points: the true points and the composite points. A composite
point consist of an infinite number of points. It is uniquely
determined by the number of unit sets. Note that, in string
theories, in order to get one natural number (mode)
one needs at least two real numbers (length, tension) and
additional assumptions.
Cardinality of the proper microscopic interval may be
regarded as some qualitative property of the point.
This property vanishes if the interval is destroyed
(decay of the corresponding point).
The minimal building block for a composite point is a unit
set. In the three-dimensional case, there must be three types of
the unit sets forming, in the macroscopic limit, three-dimensional
approximately continuous space.

\end{document}